\documentclass{vgtc}                          

\graphicspath{{figures/}{pictures/}{images/}{./}} 

\usepackage{times}                     

\usepackage{tabu}                      
\usepackage{booktabs}                  
\usepackage{lipsum}                    
\usepackage{mwe}                       
\usepackage{xcolor}
 \usepackage{amsmath}
 \usepackage{tabularx}
 \usepackage{graphicx}
\usepackage{mathptmx}                  
\usepackage{amsfonts}                  

\usepackage{siunitx}
\usepackage{makecell}
\usepackage{booktabs}  
\usepackage{multirow}

\onlineid{0}

\vgtccategory{Research}

\vgtcinsertpkg

\title{VirSqueezer: Generating Realistic Deformations and Squeezing Dynamics in VR from Fine-Grained Squeezing Controls}

\author{
        Qian Zhang\thanks{e-mail: 2330702027@st.btbu.edu.cn}\\ %
        \scriptsize School of Computer and Artificial Intelligence, \\ \scriptsize Beijing Technology and Business University, China. %
        \and Xiaoming Chen\thanks{e-mail: xiaoming.chen@btbu.edu.cn}\\ %
        \scriptsize The University of Sydney, Australia, and \\ %
        \scriptsize School of Computer and Artificial Intelligence, \\ \scriptsize Beijing Technology and Business University, China. 
        \and Xiaorui Ma\thanks{e-mail: xiaorui.ma@st.btbu.edu.cn}\\ %
        \scriptsize School of Computer and Artificial Intelligence, \\ \scriptsize Beijing Technology and Business University, China. 
         \and Haisheng Li\thanks{e-mail: lihsh@btbu.edu.cn}\\ %
        \scriptsize School of Computer and Artificial Intelligence, \\ \scriptsize Beijing Technology and Business University, China. %
        \and Weidong Cai\thanks{e-mail: tom.cai@sydney.edu.au}\\ %
        \scriptsize School of Computer Science, \\ \scriptsize The University of Sydney, Australia. }        

\teaser{
  \centering
  \includegraphics[width=1.0\linewidth]{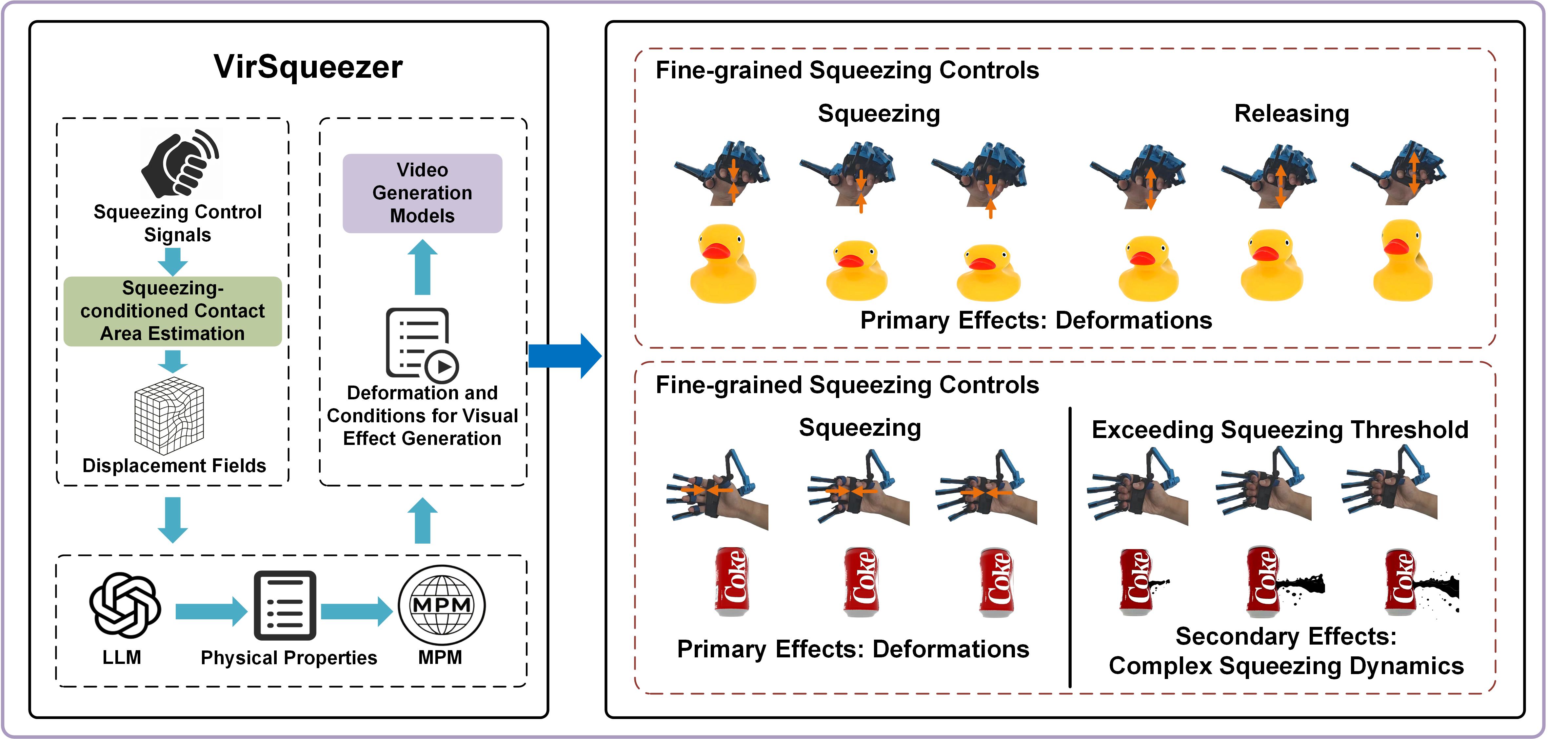}
  \caption{VirSqueezer generates localized object deformations (primary effects) and complex squeezing dynamics (secondary effects), such as rupture and overflow, driven by fine-grained, temporally evolving, finger-level squeezing controls. Through object contact area estimation, MPM-based physical simulation, and conditioned visual effect generation, VirSqueezer enables the creation of unique deformations and dynamic visual effects resulting from the user’s squeezing controls. %
  }
  \label{fig:teaser}
}

\abstract{
Squeezing is one of the most natural forms of hand manipulation, inherently involving fine-grained, temporally evolving, per-finger flexion. In VR content creation, squeezing plays a unique role in enabling particular visual effects such as localized deformations and dynamic behaviors, e.g., bursting a Coke can or juicing a fruit, thereby expanding the expressive possibilities of VR content. However, existing techniques, such as 3D Gaussian splatting-based methods and diffusion-based video generation models, are limited in their ability to simulate fine-grained virtual squeezing effects. We introduce VirSqueezer, a framework designed to generate both localized deformations (primary effects) and complex squeezing dynamics, such as rupture and overflow (secondary effects). VirSqueezer captures squeezing control signals using a SenseGlove and provides the user with inferred resistance force feedback during the squeezing process. By estimating object contact areas, inferring physical properties, and simulating physical responses, VirSqueezer computes conditions that guide generation models for visual effect generation, ensuring both visual coherence and temporal synchronization with the simulation. Consequently, VirSqueezer enables the generation of physically realistic visual effects directly from continuous, fine-grained squeezing control signals. Our extensive evaluation demonstrates VirSqueezer’s ability to reproduce realistic localized deformations, generate convincing visual dynamics, and maintain consistency in fine-grained squeezing controls.
} 


\begin{document}
\firstsection{Introduction} \label{sec:introduction}
\maketitle
Creating physically realistic immersive content in Virtual Reality (VR) often requires accurately simulating how humans interact with 3D objects, as these interactions trigger a wide range of physical phenomena and visual effects. Among these, squeezing stands out as one of the most natural forms of hand manipulation. It involves fine-grained, temporally evolving per-finger flexion, which not only induces object deformations but also triggers characteristic effects, such as rupture and overflow, as demonstrated in Figure~\ref{fig:teaser}. In VR content creation, squeezing plays a unique role in enabling realistic and engaging interactions, such as bursting a can, crushing a soft toy, or juicing fruit, thereby broadening the expressive possibilities of VR content.

However, achieving these effects remains a challenge with current VR content creation techniques. Recent advancements in VR content creation have primarily been driven by two categories of techniques: (1) 3D Gaussian Splatting (3DGS)-based frameworks ~\cite{kerbl20233d}, including Material Point Methods (MPM)-based approaches ~\cite{jiang2016mpm}, and (2) generative models. On the one hand, the introduction of 3DGS~\cite{kerbl20233d} enables the real-time rendering of photorealistic scenes with exceptional efficiency. Further developments, such as 4D Gaussians \cite{wu20244d} and VR-GS \cite{jiang2024vr}, have improved spatiotemporal coherence, enhancing the realism of VR experiences. Additionally, MPM-based methods \cite{jiang2016mpm}, which utilize Material Point Methods for physics-driven interactions, have facilitated the simulation of deformations and object manipulations for 3DGS. For example, models like PhysGaussian \cite{xie2024physgaussian} and OmniPhysGS \cite{lin2025omniphysgs} integrate MPM with 3DGS to further enhance the accuracy of physical simulations, enabling more realistic object behaviors. On the other hand, generative models have gained significant attention for their ability to generate high-fidelity, temporally consistent dynamic content. Such models like Stable Diffusion \cite{rombach2022high}, VideoCrafter \cite{wang2023videocrafter}, and AnimateDiff \cite{guo2023animatediff} can synthesize photorealistic content that evolves over time, ensuring visual consistency throughout video sequences, making them ideal for producing coherent content in VR environments. 

Despite the substantial progress of both 3DGS-based frameworks (including MPM-based methods) and generative models, each faces limitations when it comes to fine-grained content creation tasks, such as simulating the deformation and complex visual dynamics of squeezing. While 3DGS-based frameworks excel at rendering realistic object behaviors, they struggle to synchronize with fine-grained control signals like per-finger flexion and motion trajectories, which are crucial for intricate manipulations such as squeezing. Similarly, although MPM-based methods effectively capture global deformations, they face difficulties in modeling localized deformations and reproducing complex dynamics such as rupture and overflow, which are essential phenomena for realistic squeezing behaviors like a Coke can bursting or juice spilling from an orange. Generative models such as Diffusion-based models, by contrast, are proficient at generating high-quality dynamic content but often fail to maintain temporal and spatial coherence under continuous user input, particularly when handling deformations driven by fine-grained, temporally evolving, finger-level manipulations. 

To address these limitations, we propose VirSqueezer, a novel framework that leverages fine-grained, temporally evolving, finger-level control signals to drive high-fidelity physics simulations and generate physically realistic squeezing effects in VR. Using a SenseGlove ~\cite{senseglove}, VirSqueezer captures squeezing control signals, including per-finger flexion and motion trajectories, and maps them to local object deformation states through a physics modeling pipeline. Specifically, the control signals are used to estimate the object's contact area, which is then transformed into Dirichlet boundary conditions~\cite{601047}, enabling physics-based, localized deformation simulations via MPM~\cite{jiang2016mpm} and generating realistic deformation effects (primary effects). The material properties of the manipulated object are estimated using a large language model (LLM), while resistive force feedback is rendered through the SenseGlove based on these properties, creating a realistic bidirectional squeezing experience. Beyond deformation, VirSqueezer also synthesizes ``secondary effects'' of complex squeezing dynamics, such as rupture and overflow, which are essential for replicating real-world squeezing dynamics. To achieve this, VirSqueezer computes conditions based on the user’s squeezing behaviors to control generative models, including AnimateDiff~\cite{guo2023animatediff}, Stable Diffusion~\cite{rombach2022high}, and ControlNet~\cite{zhang2023adding}, ensuring that the secondary effects remain visually coherent and temporally synchronized with the simulation. 
Extensive experiments demonstrate that VirSqueezer enables a new capability for fine-grained squeezing-driven deformation and complex squeezing dynamics that is not supported by existing approaches, while achieving superior performance in visual quality, physical commonsense, and squeezing-controlled deformation consistency, as demonstrated through quantitative metrics, human evaluation, and evaluations using multiple large language models.
However, it is to be noted that the current version of VirSqueezer is not a real-time system, as the generation of complex squeezing dynamics relies on computationally intensive generative models. Achieving real-time performance while satisfying the fine-grained control and physical consistency requirements of VirSqueezer remains challenging with current generative models.

In summary, we present VirSqueezer, a framework that advances VR content creation for squeezing dynamics by bridging fine-grained finger-level controls, realistic physical simulations, and the generation of complex squeezing effects. The key contributions of this work are threefold:
\begin{itemize}
\item To the best of our knowledge, VirSqueezer is the first framework specifically designed for fine-grained virtual squeezing. It bridges continuous finger-level controls with localized physical deformation and dynamic visual effect generation, creating unique squeezing visual content in VR.
\item We introduce a fine-grained squeezing-to-deformation modeling approach that jointly considers finger-level squeezing controls, localized contact, and object material properties, enabling physically plausible localized deformation and material-aware resistive force feedback.
\item We introduce a physics-guided squeezing effect generation approach that integrates localized deformations (primary effects) with complex squeezing dynamics such as rupture and overflow (secondary effects), enabling physically realistic and temporally coherent squeezing effects.
\end{itemize}

We believe VirSqueezer lays the foundation for a new technical roadmap in physically realistic VR content creation that responds to fine-grained user controls. Our key innovation lies in the integration of Dirichlet boundary conditions with MPM to model realistic squeezing deformations, enabling more accurate material responses. Additionally, we introduce a novel control diffusion method that conditions generative models on physics-based deformation signals, ensuring that the generated dynamic visual effects are not only realistic but also consistent with the underlying physical simulations.

\section{Related Work} \label{sec:relatedwork}
Visual content creation in VR environments has long attracted broad attention from the research community. The emergence of 3DGS~\cite{kerbl20233d} and its subsequent advances have introduced an efficient and scalable explicit representation, enabling flexible content creation and editing. Moreover, generative models such as diffusion-based models have garnered significant attention for their ability to generate high-fidelity, temporally consistent dynamic content.

\subsection{3DGS-based Frameworks and MPM-based Methods}
In recent years, 3DGS~\cite{kerbl20233d} has been proposed, which fits discrete point clouds with Gaussian distributions, enabling efficient and realistic static scene reconstruction. This explicit representation has sparked significant research and advancements in content creation built upon 3DGS. For example, Chen et al. \cite{chen2024gaussianeditor} proposed GaussianEditor, a 3DGS editing method that enhances editing accuracy through Gaussian semantic tracking. Huang et al. \cite{huang2024sc} proposed SC-GS, which explicitly decomposes the motion and appearance of dynamic scenes into sparse control points and dense Gaussian functions, respectively, enabling user-controlled motion editing while preserving high-fidelity appearance. Guédon et al. \cite{guedon2024sugar} proposed SuGaR, which can extract meshes from 3DGS precisely and extremely quickly, allowing users to easily perform editing, animation, and other operations using conventional rendering engines. Jiang et al. \cite{jiang2024vr} proposed VR-GS, which develops an interactive Gaussian distribution with perceived physical dynamics within VR environments, implementing an efficient two-tier embedding strategy and deformable object simulation for real-time execution and dynamic responses.

To better simulate physical behaviors in visual content creation, an increasing number of researchers combine MPM~\cite{jiang2016mpm} with 3DGS to perform visual content creation that adheres to real-world physical laws. For instance, Xie et al. \cite{xie2024physgaussian} introduced PhysGaussian, a novel approach that seamlessly integrates physics-based Newtonian dynamics into 3D Gaussian functions, enabling high-quality, novel motion synthesis. Building on this, Huang et al. \cite{huang2025dreamphysics} proposed a method that utilizes a video generation model to generate prior knowledge of physical material fields, and then uses a physics-based MPM simulator to generate realistic four-dimensional content.
To further expand the range of material simulations in physics, Lin et al. \cite{lin2025omniphysgs} introduced a technique using pre-trained video generation models to supervise the estimation of material weight factors, which are then used in MPM for physical motion simulation. Similarly, Zhang et al. \cite{zhang2024physdreamer} proposed utilizing learned object dynamics priors from video generation models, integrating them with MPM to endow static 3D objects with interactive dynamics. Furthermore, Ni et al. \cite{ni2024phyrecon} introduced differentiable rendering and differentiable physics simulation for learning implicit surface representations, addressing physical inconsistencies in multi-view neural reconstruction. 

While the above methods provide support for accurate physical simulation of 3DGS objects, many of the physical parameters still require manual configuration. Moreover, these methods mainly focus on simulating global object motion, with a particular lack of studies on fine-grained squeezing deformations. Furthermore, these methods are unable to generate the complex visual dynamics that result from squeezing.

\subsection{Diffusion-based Models}
Compared to MPM-based methods, diffusion-based models offer enhanced scalability for content creation, but this comes at the cost of limited physical realism. Although many of these methods do not directly simulate precise physical interactions, they leverage the inherent advantages of diffusion models to simulate motion. For example, Yu et al. \cite{yu20244real} proposed a novel process for generating photo-realistic 4D scenes from text, eliminating the dependency on multi-view generation models and instead utilizing video generation models trained on diverse real-world datasets.
Zhuang et al. \cite{zhuang2024tip} introduced TIP-Editor, a 3D scene editing framework that leverages text and image prompts to guide diffusion-based generation for editing effects, while using 3D bounding boxes to define the regions to be edited. Ren et al. \cite{ren2023dreamgaussian4d} proposed DreamGaussian4D, which combines explicit spatial transformation modeling with static GS. This framework employs a diffusion model to generate object motions, providing an efficient and powerful representation for 4D generation. Wimmer et al. \cite{wimmer2025gaussians} introduced Gaussians2Life, which uses powerful video diffusion models as generative components, transforming 2D videos into meaningful 3D motion. 

There is also existing research that leverages video generation models to edit the visual effects of 3DGS objects. For example, Li et al. \cite{li2024animatable} proposed Animatable Gaussians, which leverage powerful 2D CNNs and 3D Gaussian splatting to construct highly realistic human motion models. Wu et al. \cite{wu2024gaussctrl} introduced GaussCtrl, which employs pretrained 2D diffusion models (ControlNet) conditioned on input prompts to edit 3DGS objects, thereby refining their 3D structures. Lin et al. \cite{lin2025diffsplat} presented DiffSplat, which repurposes image diffusion models to enable scalable Gaussian splat generation. Gomel et al. \cite{gomel2024diffusion} developed a novel diffusion-based framework for 3D scene editing, ensuring cross-view consistency and realism. Tang et al. \cite{tang2023dreamgaussian} proposed DreamGaussian, a generative 3D Gaussian splatting model capable of mesh extraction and texture refinement in UV space. Beyond these, some researchers have also investigated style transfer for 3DGS objects. For instance, Howil et al. \cite{howil2025clipgaussian} introduced CLIPGaussian, which supports multimodal text- and image-guided stylization across 2D images, videos, 3D objects, and 4D scenes. Liu et al. \cite{liu2025abc} proposed ABC-GS, a 3DGS-based framework that employs segmentation masks to precisely align content and style features, enabling high-quality 3D style transfer. 

However, the methods above rely solely on prompts to simulate 3DGS object motion or visual effects. They are unable to incorporate fine-grained, user-specific control signals, such as continuous squeezing, that drive localized deformations.

\section{Method} \label{sec:method}

\begin{figure*}
    \centering
    \includegraphics[width=1.0\linewidth]{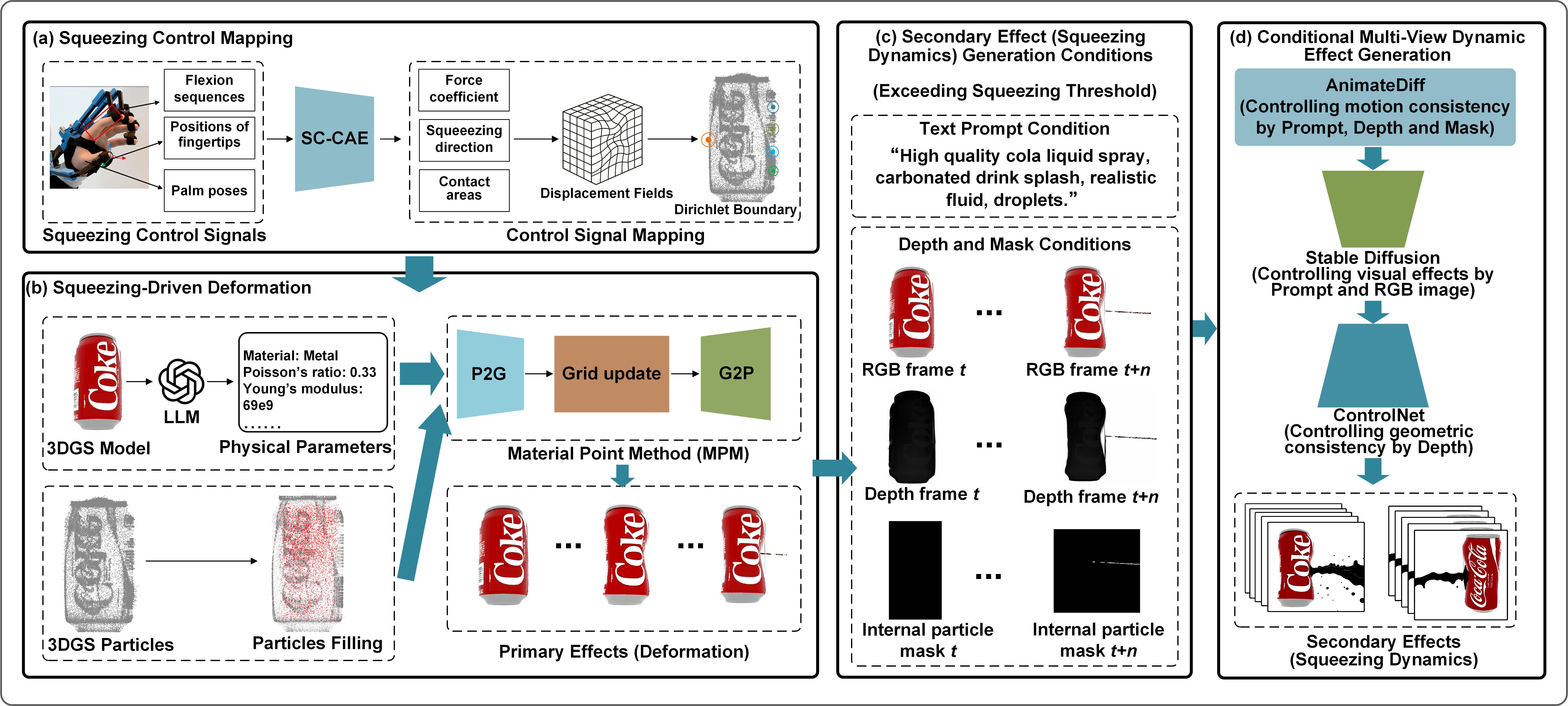}
    \caption{Structure of VirSqueezer: (a) Squeezing control signals (including flexion sequences of the five fingers, 3D positions of fingertips, and palm poses) are mapped to Dirichlet displacement constraints through squeezing-conditioned contact area estimation (SC-CAE) and finger mapping. (b) Using physical property priors inferred by the LLM, the MPM pipeline (P2G → Grid Update → G2P) generates the primary deformation effects and computes the squeezing threshold that triggers secondary effects of complex squeezing dynamics. (c) Text prompts, depth maps, and masks are computed as conditions for generating complex squeezing dynamics. (d) These generated conditions are applied to control AnimateDiff, Stable Diffusion, and ControlNet to produce realistic multi-view squeezing dynamics while maintaining motion and geometric consistency.}
    \label{fig:method}
\end{figure*}

As illustrated in Figure~\ref{fig:method}, VirSqueezer is composed of four core components:
a) Squeezing Control Mapping: The user's squeezing control signals, captured by the SenseGlove, are preprocessed and mapped to the object’s physical constraint space, creating a tight coupling between the user's squeezing and the virtual object's responses.
b) Squeezing-Driven Deformation: The MPM is employed to physically simulate the primary effects, i.e., the fine-grained squeezing deformations. This process also generates user resistive feedback based on the object's material properties during the squeezing process.
c) Secondary Effect Generation Conditions: After the squeeze deformation, we obtain the depth, RGB, and particle mask of the 3DGS object, combined with textual prompts as the conditions for generating the secondary effects of squeezing dynamics.
d) Conditional Multi-View Dynamic Effect Generation: To simulate secondary effects such as rupture and overflow, video generation models are conditioned on the user's squeezing control signals, creating visually compelling effects from multiple viewpoints for immersive VR experience.

\subsection{Squeezing Control Mapping} \label{sec:squeezing control}
As illustrated in Figure~\ref{fig:method}(a), we synchronize three primary categories of squeezing control signals along a unified timeline: 1) normalized flexion sequences of the five fingers, 2) 3D positions of individual fingertips, and 3) 6-DoF palm poses. All signals are aligned with the sampling rate of SenseGlove, and an exponential moving average is applied to suppress noise induced by hand jitter. We then adapt a squeezing-conditioned contact area estimation (SC-CAE) based on Hartmann's method \cite{hartmann2022method} to process the squeeze control signal collected offline, which first projects the squeezing control signals onto the object surface and subsequently generates a sequence of contact areas. This sequence consists of multiple localized contact regions, each corresponding to a different contact surface of the finger.
To ensure physical realism and stability, displacements on these contact areas are imposed as Dirichlet boundary conditions~\cite{601047} on the object surface, directly constraining the kinematic behavior of the contact areas. This strategy avoids the instability typically caused by directly applying forces to the object, as seen in methods like PhysGaussian~\cite{xie2024physgaussian}, and enables seamless integration with the MPM for handling deformations.

During computation, the normalized flexion sequences of the five fingers are first transformed into physically meaningful force proxies. This transformation is activated only within the effective exertion range and supports bidirectional activation to better capture different stages of squeezing behavior. In addition, nonlinear sensitivity is carefully controlled, and finger-specific baseline weights are assigned according to distinct squeezing patterns. As a result, we obtain time-varying force coefficients, which are formally defined as follows:

\begin{equation}
    c_{k,t} = w_k \left( \max \big(\tilde{u}_{k,t} - \beta, \, 0 \big) \right)^{\gamma},
\end{equation}

\noindent where $\tilde{u}_{k,t}$ denotes the de-biased flexion of finger $k$ at time $t$; $\beta$ is the activation threshold that filters out slight micro-bending; $\gamma$ is the nonlinearity index controlling the steepness of the response; $w_k$ is the finger-specific baseline weight, e.g., stronger emphasis on the thumb or index finger; and $c_{k,t}$ denotes the computed force coefficient.

After obtaining the per-finger force coefficients, we employ the aforementioned SC-CAE method to construct an appropriate displacement field on the object surface $S$. Specifically, given the predicted palm pose and fingertip positions, we treat the center of the contact area, surface normal, and anisotropic scales as unknowns, and formulate a composite objective that jointly enforces geometric compatibility, normal alignment, curvature regularization, and multi-finger force coupling. This formulation ensures that the estimated displacement field adheres to the finger-oriented contact areas, aligns correctly with the surface normals, avoids sharp high-curvature areas, and prevents overlap when multiple fingers act simultaneously. These constraints collectively guarantee spatial continuity and temporal robustness of the displacement field, thereby enabling fine-grained local deformations. The formal definition is given as follows:

\begin{equation}
\text{Displacement Field} = \min_{\mathbf{x} \in S,\, \mathbf{n},\, \sigma} \;
\begin{aligned}
& \lambda_{\text{area}} \, d^{2}\big(\mathbf{x}, \text{ray}(\mathbf{P}_{k,t}, \mathbf{a}_{k,t})\big) \\
& + \lambda_{\text{align}} \left| 1 - \mathbf{a}_{k,t}^{\top}\mathbf{n} \right| \\
& + \lambda_{\text{curvature}} \, \phi(\kappa_{1}, \kappa_{2}) \\
& + \lambda_{\text{scale}} \left\| \sigma - \sigma_{0}(\alpha_{k,t}) \right\|^{2} \\
& + \lambda_{\text{overlap}} \, \phi_{\text{overlap}}, 
\end{aligned}
\end{equation}

\noindent where \(\mathbf{x}\) denotes the center of the candidate contact area, 
\(\mathbf{n}\) is its surface normal, 
and \(\sigma\) represents the anisotropic scale, i.e., the patch radius along the principal curvature directions; 
\(\mathbf{P}_{k,t}\) and \(\mathbf{a}_{k,t}\) are the fingertip position and squeezing direction, respectively; 
\(d(\cdot,\text{ray})\) is the point-to-ray distance; 
\(\lvert 1 - \mathbf{a}_{k,t}^{\top}\mathbf{n} \rvert\) is the normal alignment term; 
\(\phi(\kappa_{1}, \kappa_{2})\) is the curvature regularization; 
\(\lVert \sigma - \sigma_{0}(\alpha_{k,t}) \rVert^{2}\) enforces the anisotropic scale to match the force prior \(\sigma_{0}(\cdot)\); 
and \(\phi_{\text{overlap}}\) penalizes multi-finger overlap; 
\(\lambda_{\text{area}}, \lambda_{\text{align}}, \lambda_{\text{curvature}}, \lambda_{\text{scale}}, \lambda_{\text{overlap}}\) are the corresponding weights.

As shown in Figure~\ref{fig:method}(a), based on the derived per-finger force coefficients and the estimated contact area, we synthesize a continuous displacement field on the surface $S$ and convert it into Dirichlet boundary constraints that can be utilized by the MPM solver. Specifically, we first apply a finger-mapping procedure to decompose the squeezing control signals into multiple finger components, displacing the contact surface along the squeezing direction. Each finger’s displacement is constrained by a global scaling factor. Finally, we combine the displacement field of each finger to construct a Dirichlet constraint, which is then provided as input to the MPM simulation. The computation is formally expressed as follows:

\begin{equation}
\Delta \mathbf{x}(\mathbf{x}) = \sum_{k} \eta \, c_{k,t} \, \phi_{k}(\mathbf{x}) \, \mathbf{n}_{k},
\end{equation}

\noindent where $\Delta \mathbf{x}(\mathbf{x})$ is the target displacement field of surface point $\mathbf{x}$; $\eta$ is a global scaling factor; $c_{k,t}$ is the force coefficient of finger $k$ at time $t$;
$\phi_{k}(\mathbf{x})$ is the anisotropic soft weight; 
and $\mathbf{n}_{k}$ denotes the patch normal, i.e., squeezing direction.

\subsection{Squeezing-Driven Deformation} \label{sec:mpm}
As illustrated in Figure~\ref{fig:method}(b), the Dirichlet boundary constraint computed in Section~\ref{sec:squeezing control} is then projected onto the grid, driving kinematic squeezing deformation of the object’s surface. This process is formulated as follows:

\begin{equation}
    \boldsymbol{\sigma} = \frac{1}{\det(\mathbf{F})} 
    \frac{\partial \Psi}{\partial \mathbf{F}_E}(\mathbf{F}_E)\mathbf{F}_E^{\top}, 
    \qquad
    \mathbf{F} = \psi(\mathbf{F}_E), 
\end{equation}

\noindent where $\boldsymbol{\sigma}$ is the Cauchy stress, $\mathbf{F}$ the total deformation gradient, $\mathbf{F}_E$ its elastic component, $\Psi$ the hyperelastic energy density, $\psi$ the material-law correction mapping, and $\det(\mathbf{F})$ is the local volume ratio \cite{feng2024pie}.

After formulating the constitutive relation and Cauchy stress update, as shown in Figure~\ref{fig:method}(b), we next discretize the dynamics for time evolution under squeezing controls by adopting the MPM ~\cite{jiang2016mpm}. Each time step follows the standard particle-to-grid (P2G), grid update, and grid-to-particle (G2P) process~\cite{jiang2016mpm}: Particle information, such as mass and momentum, is first transferred to the background grid in the P2G stage. The grid then acts as the computational grid to update nodal forces and velocities under both external squeezing inputs and internal elastic responses during the grid update stage. In the final G2P stage, the updated grid quantities are interpolated back to particles to advance their positions and deformation states. During the G2P stage, the elastic deformation is updated at each substep via the return mapping rule:

\begin{equation}
    \mathbf{F}_E^{\text{new}} = 
    \big(\mathbf{I} + \Delta t \nabla \mathbf{v}\big)\mathbf{F}_E
    \xrightarrow{\text{return}}
    \mathbf{F}_E^{\text{new}} \leftarrow \psi(\mathbf{F}_E^{\text{new}}), 
\end{equation}

\noindent where $\nabla \mathbf{v}$ denotes the local deformation trend induced by the squeezing control signals, while the ``return'' step corresponds to a material-law correction. In practice, this determines whether the object, after being squeezed, should exhibit only a primary deformation effect or also exhibit secondary effects of complex squeezing dynamics such as rupture and overflow (more on this later).

Before applying MPM to simulate object deformations, we use an LLM (GPT-5) to infer the object’s physical properties, including material type, Young’s modulus, Poisson’s ratio, and yield strength. Based on these properties, the LLM also determines a squeezing threshold for each object. When the deformation exceeds this threshold, objects with certain material properties exhibit appropriate secondary effects, such as rupture or overflow. Objects that remain below the threshold instead may produce a release effect when the squeezing control signals are relaxed. For example, a rubber duck rebounds after being released, as the applied squeezing force decreases. During the user’s squeezing control, VirSqueezer also computes the product of the material yield strength and the fingertip contact area to determine the resistive force exerted by the SenseGlove on each finger. This mechanism enables simulated resistance, establishing a bidirectional visuo-haptic loop that more closely approximates real-world squeezing behaviors.

For the generation of secondary effects for complex squeezing dynamics, we first fill the object interior with particles and initialize their position information. Unlike existing methods (such as PhysGaussian~\cite{xie2024physgaussian}) that use internal particles primarily to assist physical dynamics, we use internal particles to simulate the internal structures of objects that cannot be reconstructed by 3DGS. Both the internal particles and the 3DGS object are simulated on the same MPM grid but are assigned different material parameters. When the deformation reaches the threshold, the internal particles are released in the opposite direction of the squeezing deformation at the position where the finger contact surface is reached, with initial velocities proportional to the local finger pressure, while being subjected to viscosity, gravity, and other grid forces. However, the secondary effects generated in this manner have relatively low realism. To better optimize them with video generation models, we introduce G-buffer~\cite{chen2024gi}, which is primarily defined as a set of per-pixel scene property maps that capture the scene's geometric details and material properties.
We primarily acquire RGB images, depth maps, and mask maps containing only the motion of internal particles, which capture both the physical squeezing deformations and the motion of internal particles. These serve as conditioning inputs for subsequent visual effect generation (see next section for details).

\subsection{Conditional Multi-View Dynamic Effects Generation} 
The MPM simulation itself can hardly generate appealing secondary effects involving complex squeezing dynamics. To address this limitation, we enhance the secondary effect with the generated G-buffer conditions as described in Section \ref{sec:mpm}, including depth, RGB, and particle mask, combined with textual prompts, as shown in Figure~\ref{fig:method}(c). For viewpoint $v$ and frame $t$, we denote the corresponding G-buffer as $G_t^v=\{I_t^v,D_t^v,M_t^v\}$, where $I_t^v$, $D_t^v$, and $M_t^v$ denote the RGB image, depth map, and particle mask, respectively. 
The particle mask is first refined through connectivity checking and boundary repair, and the depth map is subsequently normalized and gated by the refined mask:

\begin{equation}
\widetilde{M}_{t}^{v}=R(M_{t}^{v}), \qquad
\widetilde{D}_{t}^{v}
=
\operatorname{Norm}(D_{t}^{v})\odot \widetilde{M}_{t}^{v},
\end{equation}
where $R(\cdot)$ denotes the mask refinement operation and 
$\odot$ denotes element-wise multiplication.
The resulting RGB frame, normalized depth map, internal particle mask, and textual prompt are then used as conditions for subsequent visual effect generation.

Subsequently, as shown in Figure~\ref{fig:method}(d), we employ the G-buffer obtained from the squeezing deformation driven by the control signals using MPM to condition the pipeline of AnimateDiff~\cite{guo2023animatediff}, Stable Diffusion~\cite{rombach2022high}, and ControlNet~\cite{zhang2023adding}.
Specifically, the prompt, depth frame, and internal particle mask are used to control AnimateDiff to establish temporal coherence in the latent space, where temporal attention enforces motion consistency across video frames.
Viewpoint-dependent depth and mask conditions are incorporated into the diffusion process, while the corresponding RGB frames and textual prompts are further used by the image-to-image module of Stable Diffusion~\cite{rombach2022high}, a latent diffusion model for high-quality image generation, for conditional generation.
Particularly, the depth condition is injected through ControlNet~\cite{zhang2023adding} to provide geometric guidance during denoising, helping preserve the spatial structure of the deformed object and maintain consistency under occlusions.
Finally, the refined foreground is blended with the background according to the particle mask, preserving the squeezing deformation while improving the visual realism of the secondary effects.

To maintain multi-view consistency with the underlying 3D representation, we perform lightweight appearance refinement on the particles obtained from Section \ref{sec:mpm}, using the refined multi-view images as reconstruction supervision.
Let $\Theta_p$ denote the optimized particle parameters, including shading parameters, opacity, and scaling in each direction.
The optimization is formulated as:

\begin{equation}
\Theta_{p}^{*}
=
\arg\min_{\Theta_{p}}
\sum_{v,t}
L_{\mathrm{rec}}
\left(
R(\Theta_{p};v,t),
I_{t}^{v,*}
\right),
\tag{7}
\end{equation}

\noindent where $R(\Theta_p;v,t)$ denotes the rendered result at viewpoint $v$ and frame $t$, 
$I_t^{v,*}$ denotes the corresponding refined image, and 
$L_{\mathrm{rec}}$ denotes the multi-view reconstruction loss.
The Gaussian distributions of solid objects and background remain frozen during this process to avoid instabilities.
This process optimizes the generated images into a realistic 4D output while supporting multi-view rendering, allowing the secondary effects to remain spatially consistent with the underlying scene structure and evolve smoothly across frames.
Additional technical details, hyperparameter settings, and reproducibility-related configurations are provided in \textbf{Part 1 of Supplementary Materials}.

\section{Evaluation} \label{sec:evaluation}
We conduct extensive experiments to evaluate the ability of VirSqueezer to generate squeezing-driven, physically plausible visual effects. We first report quantitative comparisons against MPM-based and diffusion-based baselines, measuring both visual quality and physical realism. We then conduct a user study (as ``human evaluation'') and also complement it with assessments from multiple LLM models. In addition, drawing on prior work, we introduce a set of custom metrics tailored to the evaluation on the squeezing controlled deformation and benchmark them against the baselines. Subsequently, we conduct ablation studies to investigate the contribution of key components in generating primary and secondary effects. Beyond quantitative results, we also present qualitative results of primary squeeze deformations and secondary effects such as rupture and overflow. VirSqueezer is implemented in PyTorch and the implementation details are also provided in \textbf{Part 1 of Supplementary Materials}. 

\subsection{Quantitative Evaluation} \label{sec:quantitative}
Our quantitative evaluation covers the following five aspects. We employ VBench~\cite{huang2024vbench} to assess imaging quality and motion smoothness, and physical commonsense (PC) scores~\cite{bansal2025videophy} for assessing physical realism. Additionally, following PhysDreamer~\cite{zhang2024physdreamer}, we conduct a user study (referred to as ``Human Evaluation'' as in \cite{chen2025physgen3d}) as well as evaluations using multiple LLMs~\cite{chen2025physgen3d, zhang2024physdreamer} to assess visual quality and motion realism. Moreover, we introduce four task-specific metrics tailored for our task to evaluate the squeezing controlled deformation (detailed in the subsequent sections). 

For the selection of objects in the evaluation, we followed OmniPhysGS~\cite{lin2025omniphysgs} and further expanded the evaluation to eight objects representing diverse real-world materials and squeezing behaviors. All eight objects were generated using BlenderNeRF~\cite{BlenderNeRF}, including the Can, Plush toy, Rubber duck, Clay, Tomato, Orange, Rubber ball, and Paper cup. These objects span diverse material and structural characteristics, including elastic, plastic, soft and compressible, fluid-filled, and thin-shell objects. They also exhibit a broad range of squeezing responses, including elastic deformation and recovery, plastic and non-recoverable deformation, soft compression, thin-shell buckling and collapse, rupture, internal-material extrusion, and fluid overflow.

\begin{table*}[h]
\centering
\caption{
Quantitative comparison of VirSqueezer with MPM-based methods and diffusion-based models. VirSqueezer achieves the best overall performance across physical commonsense, VBench, human evaluations, and LLM evaluations.
}
\renewcommand{\arraystretch}{1.15}
\setlength{\tabcolsep}{4.2pt}
\small
\resizebox{\textwidth}{!}{%
\begin{tabular}{|c|c|cc|ccc|ccc|}
\hline
\multirow{2}{*}{\textbf{Method}} &
\multirow{2}{*}{\centering \makecell{\textbf{Physical} \\ \textbf{Commonsense \cite{bansal2025videophy}} $\uparrow$}} &
\multicolumn{2}{c|}{\textbf{VBench \cite{huang2024vbench}}} &
\multicolumn{3}{c|}{\textbf{Human Evaluation \cite{zhang2024physdreamer}}} &
\multicolumn{3}{c|}{
\makecell{
\textbf{GPT-5 Evaluation \cite{chen2025physgen3d,zhang2024physdreamer}}\\
\fontsize{6}{7}{(\textbf{Results for more LLMs in Supplementary Material})}
}
} \\
\cline{3-4}\cline{5-10}
& &
\makecell{\textbf{Imaging Quality}~$\uparrow$} &
\makecell{\textbf{Motion Smoothness}~$\uparrow$} &
\makecell{\textbf{Visual}\\\hspace{0.25em}\textbf{ Quality}~$\uparrow$} &
\makecell{\textbf{Motion}\\\hspace{0.25em}\textbf{ Realism}~$\uparrow$} &
\makecell{\textbf{Motion}\\\hspace{0.25em}\textbf{ Consistency}~$\uparrow$} &
\makecell{\textbf{Visual}\\\hspace{0.25em}\textbf{ Quality}~$\uparrow$} &
\makecell{\textbf{Motion}\\\hspace{0.25em}\textbf{ Realism}~$\uparrow$} &
\makecell{\textbf{Motion}\\\hspace{0.25em}\textbf{ Consistency}~$\uparrow$} \\
\hline
PhysGaussian \cite{xie2024physgaussian}        & 0.354 & 0.522 & 0.994 & 3.2 & 3.1 & 3.0 & \textbf{4.0} & 3.8 & 4.1 \\
OmniPhysGS \cite{lin2025omniphysgs}           & 0.345 & 0.516 & 0.994 & 2.6 & 2.6 & 2.8 & 3.8 & 3.9 & 3.7 \\
DreamGaussian4D \cite{ren2023dreamgaussian4d} & 0.071 & 0.218 & 0.983 & 2.0 & 2.3 & 2.1 & 3.1 & 3.5 & 3.1 \\
Gaussians-to-Life \cite{wimmer2025gaussians}  & 0.234 & 0.556 & 0.987 & 2.2 & 2.3 & 2.7 & 3.7 & 3.5 & 3.2 \\
\textbf{VirSqueezer (Ours)}                                & \textbf{0.368} & \textbf{0.596} & \textbf{0.996} & \textbf{3.9} & \textbf{4.1} & \textbf{4.2} & 3.9 & \textbf{4.2} & \textbf{4.3} \\
\hline
\end{tabular}
}
\label{tab:Quantitative_1}
\end{table*}

\subsubsection{VBench Evaluation}
We use the VBench~\cite{huang2024vbench} metric to assess imaging quality and motion smoothness. VBench is a comprehensive, hierarchical evaluation tool that decomposes video generation quality into multiple well-defined dimensions, allowing for fine-grained and objective assessment. VBench consists of 16 dimensions in total. Each dimension is supported by human preference annotations, providing insights from various perspectives. Following PhysGen3D~\cite{chen2025physgen3d}, we adopt the two most relevant dimensions, imaging quality and motion smoothness, to evaluate VirSqueezer and all baselines.
For baselines, we compare VirSqueezer against state-of-the-art MPM-based methods, including PhysGaussian~\cite{xie2024physgaussian} and OmniPhysGS~\cite{lin2025omniphysgs}, as well as diffusion-based approaches, including DreamGaussian4D~\cite{ren2023dreamgaussian4d} and Gaussians-to-Life~\cite{wimmer2025gaussians}. 
As shown in Table~\ref{tab:Quantitative_1}, VirSqueezer outperforms all the baseline methods in both imaging quality and motion smoothness. These results, however, should be interpreted as demonstrating the capability and suitability of VirSqueezer for dynamic generation driven by fine-grained squeezing controls, rather than as demonstrating general superiority in video generation capability.

\subsubsection{Physical Commonsense Evaluation}
We further conduct a quantitative comparison using the Physical Commonsense (PC) metric from VideoPhy-2~\cite{bansal2025videophy}. VideoPhy-2 is a benchmark designed to assess whether generated videos align with everyday physical commonsense across a variety of activities. Its reliability is validated through extensive human judgments and is complemented by the automatic evaluator, VideoCon-Physics~\cite{bansal2024videophy}, for large-scale assessments. As shown in Table~\ref{tab:Quantitative_1}, VirSqueezer achieves the highest PC score among MPM-based methods and, by a significant margin, outperforms diffusion-based baselines, demonstrating strong alignment with physical commonsense.

\subsubsection{Human Evaluation}
Following PhysGen3D~\cite{liu2024physics3d} and PhysDreamer~\cite{zhang2024physdreamer}, we conducted a user study as ``Human Evaluation''.  This study involved 50 participants, including 27 males and 23 females, with a mean age of 22 years and an SD of 2.3 (details on the participant recruitment process are provided in \textbf{Part 2 of Supplementary Materials}). Each participant wore a SenseGlove (Development Kit) to collect squeezing control signals for multiple objects. The collected control signals were adapted in two ways: (1) as per-finger flexion and motion trajectories for an MPM-based model, and (2) as text prompts for a diffusion-based model. This enabled each baseline to generate corresponding squeezing deformation effects. Participants viewed the generated visual results in randomized order using an Oculus Quest 2 and completed a questionnaire to assess visual quality and motion realism on a 5-point Likert scale (1 = worst, 5 = best). In addition to scoring visual quality and motion realism, we also introduced a new dimension, motion consistency, to measure the alignment between a participant’s squeezing behaviors and the resulting visual effects.
During the squeezing control, we use the material yield strength estimated by the LLM (Section~\ref{sec:mpm}) to set the resistance feedback rendered through the SenseGlove, thereby creating a realistic bidirectional squeezing experience for users. As shown in Table~\ref{tab:Quantitative_1}, VirSqueezer consistently received higher ratings, with notable improvements in visual quality, motion realism, and motion consistency. 

\subsubsection{Evaluation with Multiple LLMs}
To complement the ``Human Evaluation'', following PhysGen3D~\cite{liu2024physics3d} and PhysDreamer~\cite{zhang2024physdreamer}, we further evaluate visual quality, motion realism, and motion consistency using multiple LLMs, including GPT-5, Gemini-3.1, Grok-4.1, and Manus-1.6. The use of multiple LLM evaluators provides complementary evidence to the human assessments while reducing potential model-specific bias associated with relying on a single LLM. As done in PhysGen3D~\cite{liu2024physics3d} and PhysDreamer~\cite{zhang2024physdreamer}, we convert the squeezing control signals into visualization videos and submit them, together with the corresponding squeezing effects generated by each model, for scoring on a five-point Likert scale (1 = worst, 5 = best). Due to space limitations, Table~\ref{tab:Quantitative_1} reports the results obtained using GPT-5, while the full results from all the four LLMs are provided in \textbf{Part 3 of Supplementary Materials}. As shown in Table~\ref{tab:Quantitative_1}, although VirSqueezer achieves slightly lower visual quality than PhysGaussian~\cite{xie2024physgaussian}, it significantly outperforms all other baseline models in visual quality and achieves superior performance in motion realism and motion consistency.

\subsubsection{Evaluation for Squeezing Controlled Deformation} \label{sec:Evaluation Metrics for Squeezing Interaction}
To better evaluate the effectiveness and generalizability of VirSqueezer in handling squeezing controls, we also designed tailored quantitative metrics, including temporal consistency of boundary deformation (BDC-T), spatial consistency of boundary deformation (BDC-S), squeezing-deformation correlation (SDC), and also employed the mIoU metric in our context.

We first refer to D3TW~\cite{chang2019d3tw} for addressing weakly supervised motion alignment and segmentation in videos and MLS-MPM~\cite{hu2018moving} for simulating displacement discontinuities and bidirectional rigid-body coupled motion, and define boundary deformation consistency (BDC) to verify the fidelity of squeeze deformations driven by Dirichlet boundary constraints, i.e., whether the squeezing deformation effects generated by VirSqueezer remain consistent in time and space with the displacement fields generated from the squeezing control signals. We evaluate this using two sub-metrics, BDC-T and BDC-S. BDC-T measures the temporal consistency between the specified boundary displacements and the observed deformation sequence, while BDC-S follows the mIoU paradigm and quantifies the spatial overlap between the finger-oriented contact areas and the regions of large deformation. Scores for both metrics are normalized to the range [0,1], with higher values indicating better performance. We further refer to VBench~\cite{huang2024vbench} and D3TW~\cite{chang2019d3tw} to define squeezing-deformation correlation (SDC), which evaluates the degree of temporal coupling between the captured squeezing control signals and the deformation effects generated by VirSqueezer, i.e., whether the squeezing deformation process produced by VirSqueezer aligns with the temporal evolution of the collected squeezing control signals. 
Following ContactPose~\cite{brahmbhatt2020contactpose} and S²Contact~\cite{tse2022s}, we use mIoU to verify whether deformations occur in the correct locations, i.e., whether the regions of squeezing deformation generated by VirSqueezer overlap with the regions inferred from the squeezing control signals. Additional physical quantity analysis, including finger-wise force coefficients, deformation magnitude, and displacement-field evolution, are provided in \textbf{Part 4 of Supplementary Materials}.

\subsection{Ablation Study}
\subsubsection{Effect of Dirichlet Boundary Constraints}

We conduct an ablation study to evaluate the contribution of Dirichlet boundary constraints to squeezing-driven deformation. Since PhysGaussian~\cite{xie2024physgaussian} also employs an MPM-based force-driven deformation simulation but does not incorporate explicit boundary constraints, it provides a suitable basis for isolating the contribution of Dirichlet boundary constraints. Therefore, we construct an ablated variant of VirSqueezer, denoted ``VirSqueezer w/o Dirichlet Boundary Constraint'' (or ``PhysGaussian+''), by removing the Dirichlet boundary construction and adopting a direct force-driven deformation strategy similar to PhysGaussian. In the ablated variant, we bypass the conversion of squeezing control signals into Dirichlet boundary conditions~\cite{601047} and instead directly estimate the applied force from the time-varying finger flexion, following the force-driven strategy of PhysGaussian~\cite{xie2024physgaussian}. The estimated force is then mapped onto the object surface according to the approximated contact areas and squeezing directions, and the subsequent deformation is simulated using the pulse mode of MPM.
We then compare VirSqueezer w/o Dirichlet Boundary Constraint with the full VirSqueezer using the BDC-T, BDC-S, SDC, and mIoU metrics as described in Section \ref{sec:Evaluation Metrics for Squeezing Interaction}. As shown in Table~\ref{tab:Quantitative_2}, the full VirSqueezer achieves higher average BDC-T, SDC, and mIoU scores while maintaining comparable BDC-S performance, demonstrating the overall effectiveness of constructing Dirichlet boundary displacements from squeezing control signals.

\newcolumntype{Y}{>{\centering\arraybackslash}X} 
\begin{table*}[ht!]
\centering
\caption{
Ablation study of Dirichlet boundary constraints. We compare the full VirSqueezer with an ablated variant, VirSqueezer w/o Dirichlet Boundary Constraint, using the BDC-T, BDC-S, SDC, and mIoU metrics.
}
\setlength{\tabcolsep}{4pt}
\small
\renewcommand{\arraystretch}{1.10} 
\begin{tabular}{|c|c|*{8}{c|}c|}
\hline
\multirow{2}{*}{\textbf{Method}} & \multirow{2}{*}{\textbf{Metric}} & \multicolumn{8}{c|}{\textbf{Objects}} & \multirow{2}{*}{\textbf{Avg.}} \\ \cline{3-10} 
& & \textbf{Can} & \textbf{Plush toy} & \textbf{Tomato} & \textbf{Rubber duck} & \textbf{Clay} & \textbf{Orange} & \textbf{Rubber ball} & \textbf{Paper cup} & \\ 
\hline
\multirow{4}{*}{VirSqueezer w/o Dirichlet Boundary Constraint} 
  & BDC-T &0.85  &0.77  &0.73  &0.82  &0.73  &0.69  &0.72  &0.77  &0.764  \\
  & BDC-S &0.69  &0.67  &0.63  &0.75  &0.60  &0.55  &0.66  &0.68  &\textbf{0.654}  \\
  & SDC   &0.82  &0.79  &0.70  &0.89  &0.77  &0.64  &0.77  &0.79  &0.771  \\
  & mIoU  &0.72  &0.63  &0.70  &0.67  &0.65  &0.62  &0.64  &0.66  &0.661  \\
\hline
\multirow{4}{*}{\textbf{Full VirSqueezer}} 
  & BDC-T & 0.86 & 0.75 & 0.77 & 0.85 & 0.72 & 0.69 & 0.77 & 0.79 &\textbf{0.775}  \\
  & BDC-S & 0.69 & 0.65 & 0.62 & 0.72 & 0.61 & 0.57 & 0.67 & 0.65 &0.648  \\
  & SDC   & 0.87 & 0.79 & 0.72 & 0.83 & 0.79 & 0.69 & 0.78 & 0.75 &\textbf{0.778}  \\
  & mIoU  & 0.70 & 0.67 & 0.69 & 0.70 & 0.65 & 0.65 & 0.67 & 0.62 &\textbf{0.669}  \\
\hline
\end{tabular}
\label{tab:Quantitative_2}
\end{table*}

To further analyze the contribution of Dirichlet boundary constraints, we also compare the visual results of full VirSqueezer with VirSqueezer w/o Dirichlet Boundary Constraint. The results are presented in \textbf{Part 5 of Supplementary Materials}. 
Moreover, to quantify the temporal and spatial alignment between sensed hand motion and reconstructed deformation trajectories, we provide additional evaluation with trajectory-level metrics in \textbf{Part 6 of Supplementary Materials}, including ATE~\cite{Kayan_2025_ICCV}, MTE~\cite{liu2024frechetvideomotiondistance}, Time Consistency Error, and Cosine Similarity.

\subsubsection{Effect of Physics-guided Conditions}

\begin{table*}[t]
\caption{
Ablation study of physics-guided conditions for secondary effects generation. We compare the full VirSqueezer with an ablated variant, VirSqueezer w/o Physics-guided Conditions, using the Physical Commonsense, Motion Smoothness, Motion Realism, and, Motion Consistency metrics.
}
\centering
\setlength{\tabcolsep}{4pt}
\small
\renewcommand{\arraystretch}{1.10} 

\begin{tabular}{|c|c|*{8}{c|}c|}
\hline
\multirow{2}{*}{\textbf{Method}} 
& \multirow{2}{*}{\textbf{Metric}} 
& \multicolumn{8}{c|}{\textbf{Objects}} 
& \multirow{2}{*}{\textbf{Avg.}} \\ 
\cline{3-10}

& 
& \textbf{Can} 
& \textbf{Plush toy} 
& \textbf{Tomato} 
& \textbf{Rubber duck} 
& \textbf{Clay} 
& \textbf{Orange} 
& \textbf{Rubber ball} 
& \textbf{Paper cup} 
& \\ 
\hline

\multirow{4}{*}{\makecell{VirSqueezer w/o\\Physics-guided Conditions}} 
  & Physical Commonsense 
  & 0.301 & 0.284 & 0.295 & 0.310 & 0.276 & 0.291 & 0.303 & 0.287 & 0.293 \\

  & Motion Smoothness 
  & 0.542 & 0.531 & 0.548 & 0.556 & 0.520 & 0.535 & 0.544 & 0.526 & 0.538 \\

  & Motion Realism 
  & 2.9 & 2.8 & 3.0 & 3.1 & 2.7 & 2.9 & 3.0 & 2.8 & 2.90 \\

  & Motion Consistency 
  & 2.7 & 2.6 & 2.8 & 2.9 & 2.5 & 2.7 & 2.8 & 2.6 & 2.70 \\
\hline

\multirow{4}{*}{\textbf{Full VirSqueezer}} 
  & Physical Commonsense 
  & 0.368 & 0.361 & 0.372 & 0.375 & 0.354 & 0.369 & 0.371 & 0.358 & \textbf{0.366} \\

  & Motion Smoothness 
  & 0.596 & 0.588 & 0.602 & 0.607 & 0.581 & 0.594 & 0.601 & 0.587 & \textbf{0.595} \\

  & Motion Realism 
  & 4.1 & 4.0 & 4.2 & 4.3 & 3.9 & 4.1 & 4.2 & 4.0 & \textbf{4.10} \\

  & Motion Consistency 
  & 4.3 & 4.2 & 4.4 & 4.5 & 4.1 & 4.3 & 4.4 & 4.2 & \textbf{4.30} \\
\hline

\end{tabular}
\label{tab:physics_condition_ablation}
\end{table*}

To investigate the contribution of physics-guided conditions in generating complex squeezing dynamics, we conduct another ablation study by removing the G-buffer conditions derived from the MPM simulation. Specifically, we construct an ablated variant, denoted as VirSqueezer w/o Physics-guided Conditions, where only textual prompts are provided to the generative models for secondary effect generation, without using the RGB frames, depth maps, and internal particle masks generated from the physical simulation.
As shown in Table~\ref{tab:physics_condition_ablation}, removing the physics-guided conditions leads to performance degradation in Physical Commonsense, Motion Smoothness, Motion Realism, and Motion Consistency. This indicates that the G-buffer conditions effectively constrain the generated secondary effects with the underlying physical deformation, improving both temporal coherence and physical plausibility. Additional qualitative comparisons of secondary effects with and without physics-guided conditions are provided in \textbf{Part 5 of Supplementary Materials}.

\subsection{Qualitative Results}
Our qualitative results encompass the following key components: (1) demonstrating the deformation effects and complex squeezing dynamics of VirSqueezer across diverse objects, (2) qualitatively comparing the overall effects of VirSqueezer against other baseline models, and (3) specifically evaluating the secondary effects through qualitative comparisons with state-of-the-art 2D video generation models. 

\subsubsection{Demonstration of Deformation Effects and Complex Squeezing Dynamics}
\begin{figure*}[h!]
    \centering
    \includegraphics[width=.93\linewidth]{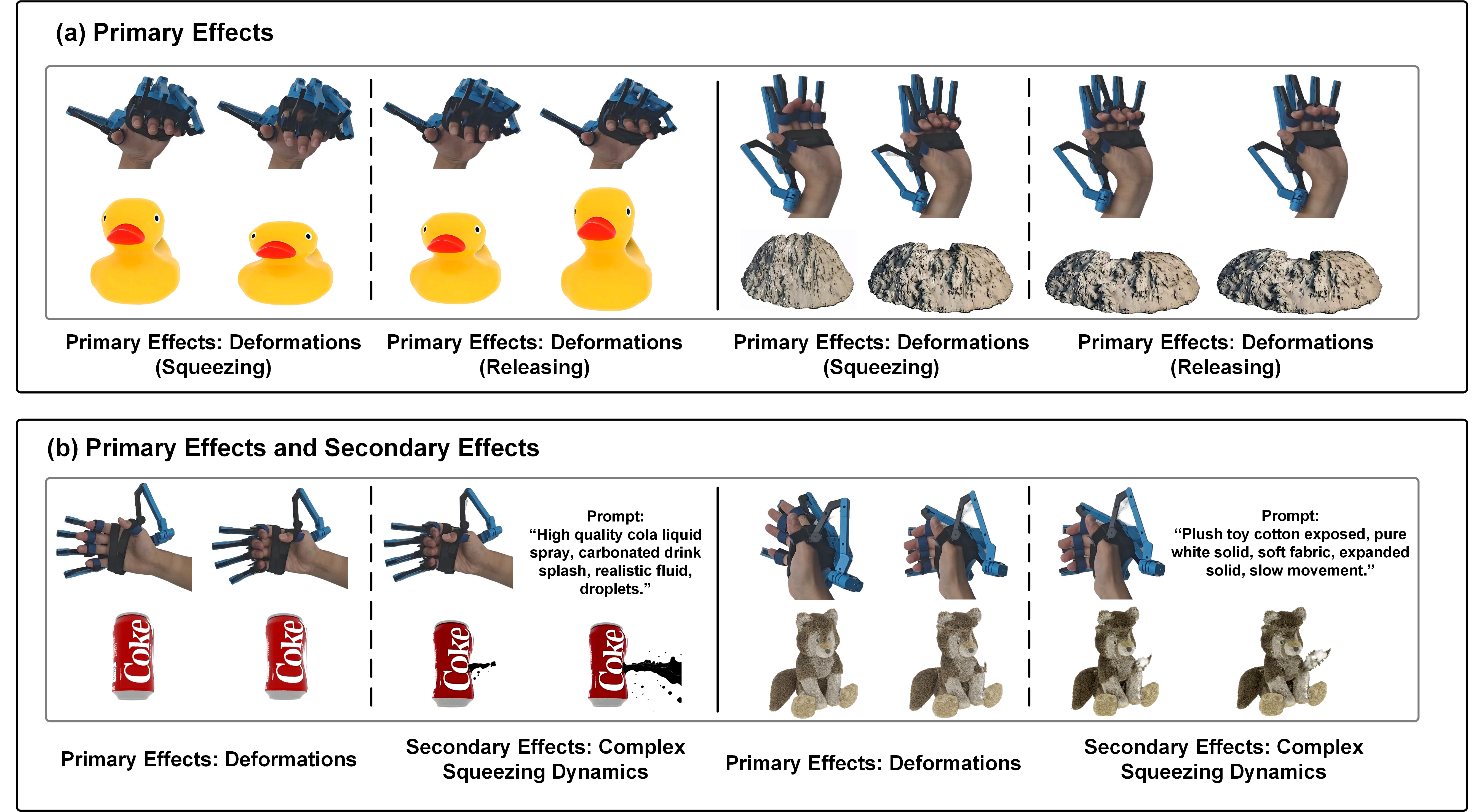}
    \caption{Demonstration of qualitative results by VirSqueezer: (a) Primary effects (deformations); 
    (b) Primary effects (deformations) and secondary effects (complex squeezing dynamics).}
    \label{fig:Qualitative}
\end{figure*}

We showcase the deformation effects of VirSqueezer when interacting with various objects, as illustrated in Figure~\ref{fig:Qualitative}. The objects experimented with are the same as described in Section~\ref{sec:quantitative}.
In Figure~\ref{fig:Qualitative}(a), the original objects undergo the primary effects of deformation in response to the user’s squeezing controls. As the squeezing force is gradually released, the objects exhibit corresponding release behaviors, such as rebound or non-rebound responses. In contrast, Figure~\ref{fig:Qualitative}(b) illustrates cases where the deformation exceeds the squeezing threshold defined in Section~\ref{sec:mpm}, leading to the emergence of secondary effects of complex squeezing dynamics. For example, when a soda can is squeezed beyond the threshold, cola is ejected from the deformed can,
and when a plush toy is squeezed beyond the threshold, cotton stuffing spills from its interior. Additional qualitative results covering primary deformations and complex squeezing dynamics across different objects are provided in \textbf{Part 5 of Supplementary Materials}.

\subsubsection{Overall Comparison on Deformation Effects and Complex Squeezing Dynamics}

\begin{figure}[h]
    \centering
    \includegraphics[width=\linewidth]{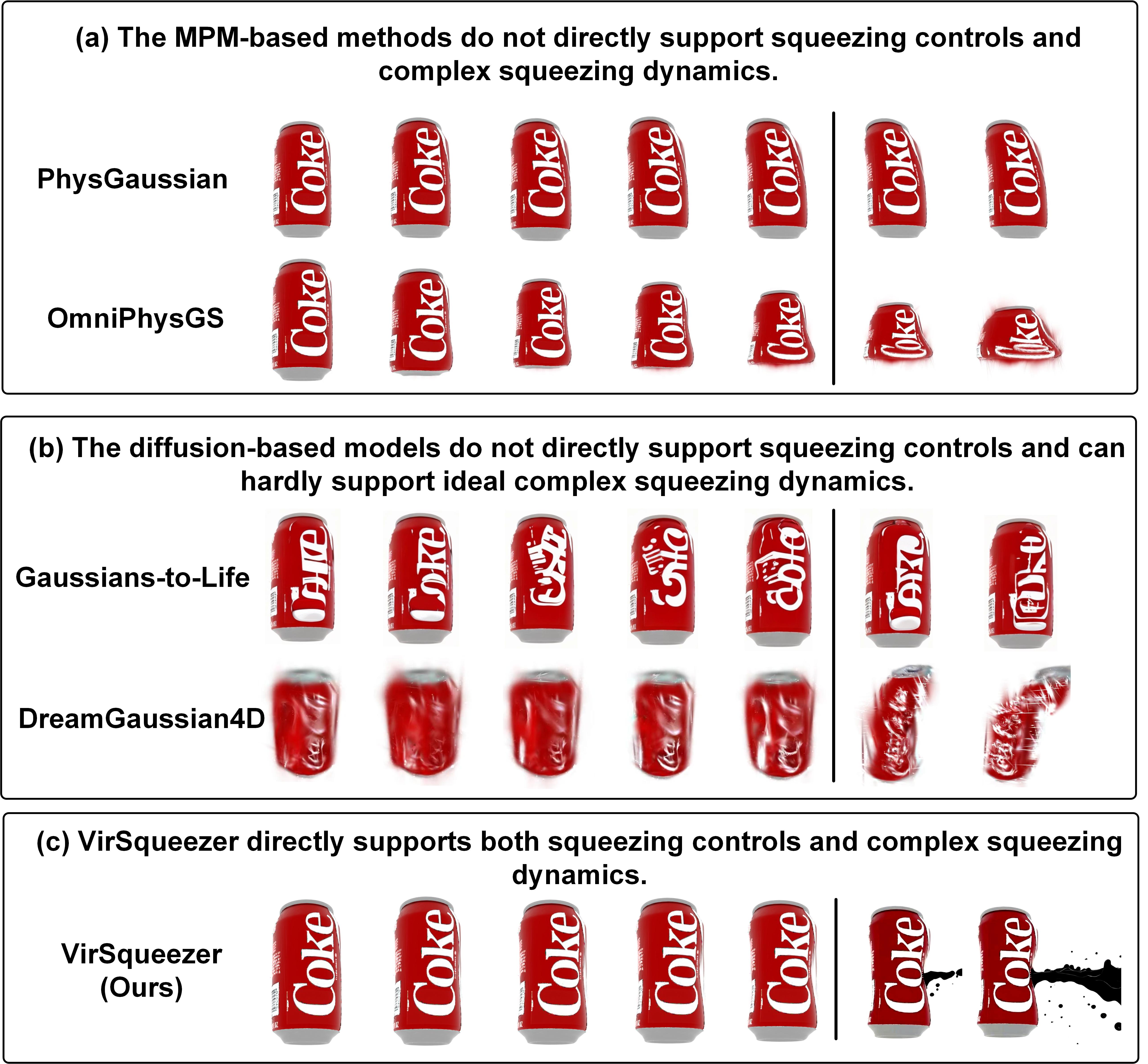}
    \caption{
    Overall qualitative comparison of primary effects (deformations) and secondary effects (complex squeezing dynamics) generated by VirSqueezer, MPM-based methods, and diffusion-based models. The results show that VirSqueezer more accurately follows the fine-grained squeezing controls while generating physically plausible deformations and realistic complex squeezing dynamics.
    }
    \label{fig:Qualitative_compare}
\end{figure}

In Figure~\ref{fig:Qualitative_compare}, we provide qualitative comparisons of VirSqueezer with MPM-based models, including PhysGaussian~\cite{xie2024physgaussian} and OmniPhysGS~\cite{lin2025omniphysgs}, as well as diffusion-based models, including Gaussians-to-Life~\cite{wimmer2025gaussians} and DreamGaussian4D~\cite{ren2023dreamgaussian4d}, using the same set of squeezing controls. Since these baseline methods do not directly support fine-grained squeezing control signals, we adapt their inputs according to their respective modeling paradigms. For the MPM-based methods, the applied force is estimated from the temporal evolution of finger flexion and mapped onto corresponding force regions on the object surface for physics-based simulation. For the diffusion-based methods, the squeezing control signals are converted into textual prompts to guide dynamic generation.

As shown in Figure~\ref{fig:Qualitative_compare}(a), the MPM-based methods can generate physical deformations, but they do not accurately follow the intended fine-grained squeezing controls and cannot reproduce complex squeezing dynamics such as rupture and overflow. As shown in Figure~\ref{fig:Qualitative_compare}(b), the diffusion-based methods can synthesize dynamic visual effects, but their generated motions and geometries are less consistent with the squeezing controls and the underlying physical deformation. In contrast, as shown in Figure~\ref{fig:Qualitative_compare}(c), VirSqueezer directly maps fine-grained squeezing controls to localized physical deformations while further generating realistic complex squeezing dynamics, thereby maintaining both deformation fidelity and visual realism.

\subsubsection{Extended Comparison on Complex Squeezing Dynamics}
\begin{figure}[h]
    \centering
    \includegraphics[width=\linewidth]{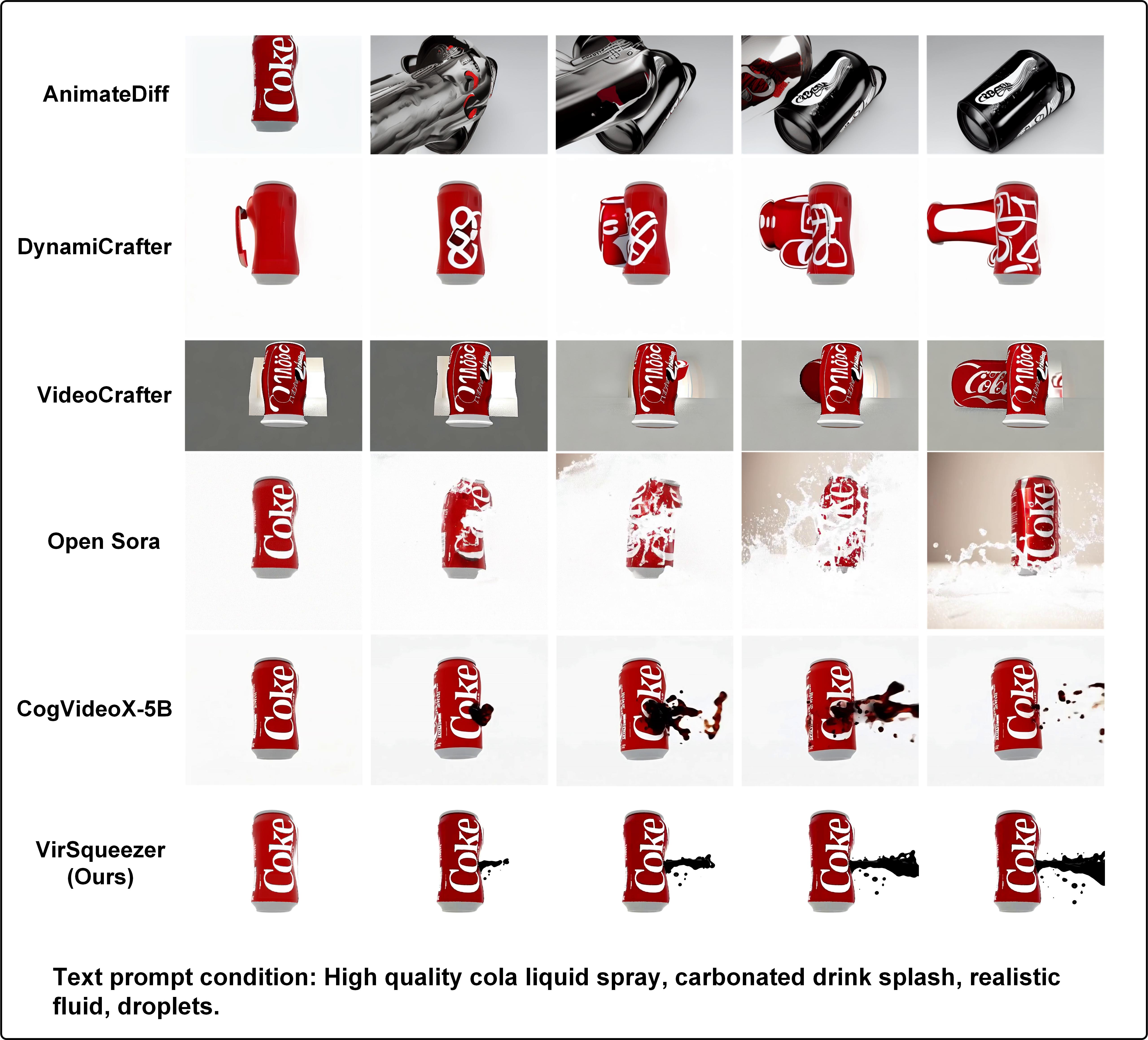}
    \caption{Extended qualitative comparison on Complex Squeezing Dynamics by VirSqueezer with baseline methods. The results demonstrate that VirSqueezer generates visually accurate squeezing dynamics while preserving the integrity of the original squeezing deformation, delivering realistic and complex squeezing dynamics.}
    \label{fig:Qualitative_compare_2}
\end{figure}

As described in Section~\ref{sec:mpm}, the complex squeezing dynamics by VirSqueezer are generated using conditional AnimateDiff, Stable Diffusion, and ControlNet, based on the G-buffer conditions (depth map, RGB frame, and internal particle mask) produced during the squeezing deformation simulation. 
To further validate the necessity and effectiveness of this design, we extend our evaluation on complex squeezing dynamics by qualitatively comparing the generated visual results with those from other video generation models using the same viewpoints and prompts, including AnimateDiff~\cite{guo2023animatediff}, DynamiCrafter~\cite{xing2024dynamicrafter}, VideoCrafter~\cite{wang2023videocrafter}, Open-Sora~\cite{zheng2024open}, and CogVideoX-5B~\cite{yang2024cogvideox}. In all cases, we emphasize in the prompts not to alter the geometry of the squeezed objects. As shown in Figure~\ref{fig:Qualitative_compare_2}, models such as AnimateDiff, DynamiCrafter, and VideoCrafter struggle to accurately generate the expected squeezing dynamics, while CogVideoX-5B and Open Sora, although partially successful, still modify the geometry of the squeezed objects, failing to maintain geometric consistency. VirSqueezer, however, preserves the original squeezing deformation while generating visually appealing squeezing dynamics, demonstrating the effectiveness of our design.

\section{Limitation} \label{sec:limitation}
While VirSqueezer advances VR content creation for fine-grained deformation and complex squeezing dynamics, several limitations remain and provide directions for future research. 

First, the generation of complex squeezing dynamics relies on computationally intensive generative models. Achieving real-time generation for VirSqueezer remains beyond the practical capabilities of current generative models. As more efficient generative models are expected to become available, their incorporation into VirSqueezer could be explored to support real-time squeezing-driven generation. In parallel, we plan to investigate Graph Neural Networks (GNNs) for squeezing motion prediction, potentially enabling real-time interactions while preserving visual fidelity and physical plausibility.

Second, the physical properties and squeezing thresholds inferred by the LLM are based primarily on general material knowledge and may not accurately characterize specific objects. Future work could incorporate measured material properties or learned physical priors to improve simulation fidelity. For example, inspired by PhysDreamer~\cite{zhang2024physdreamer}, we plan to leverage physical dynamics priors learned from video generation models and combine them with LLM-based inference to estimate more accurate and comprehensive physical parameters for 3DGS objects.

Third, the current framework and evaluation primarily focus on single-object squeezing and a predefined range of deformation and secondary effects. Extending VirSqueezer to multi-object scenarios, more diverse material behaviors, and broader hand-object manipulations would further improve and evaluate its generalizability to complex VR interactions. For example, we plan to extend the framework to multi-object settings by drawing on Feature 3DGS~\cite{zhou2024feature} and PhysGen3D~\cite{chen2025physgen3d}, thereby supporting more complex VR environments.

At last, while our evaluation covers representative objects spanning diverse materials and squeezing behaviors, broader experimental validation would further strengthen the evaluation of VirSqueezer. Future work could expand the evaluation to more object categories, material types, squeezing behaviors, and participants to more comprehensively assess its robustness and generalizability.

\section{Conclusion} \label{sec:conclusion}
We introduce VirSqueezer, a novel framework for generating realistic squeezing deformations and complex squeezing dynamics of 3DGS objects in VR, driven directly by fine-grained, temporally evolving, finger-level control signals. Using SenseGlove, we capture normalized finger flexion sequences, fingertip positions, and palm poses, which are mapped onto 3DGS object surfaces to construct Dirichlet boundary conditions. These boundary conditions enable physics-based deformation simulation via MPM, producing realistic primary effects of squeezing. To enhance visual fidelity, VirSqueezer leverages user squeezing controls to condition generative models, synthesizing secondary effects such as rupture and overflow, thereby creating a more immersive VR experience.

Despite the limitations discussed in Section \ref{sec:limitation}, VirSqueezer demonstrates a new capability for creating fine-grained squeezing-driven deformation and complex squeezing dynamics in VR. By bridging continuous finger-level controls with physically grounded deformation and generative dynamic effects, VirSqueezer extends VR content creation beyond predefined object animations and conventional deformation simulation toward more expressive, user-driven generation of dynamic virtual content.

\label{sec:figure_credits}

\bibliographystyle{abbrv-doi}

\bibliography{template}

@article{wang2023videocrafter,
  title={VideoCrafter: Open diffusion models for high-quality video generation},
  author={Wang, Yujun and Gu, Shixiang and others},
  journal={arXiv preprint arXiv:2310.19512},
  year={2023}
}

@article{jiang2016mpm,
  title={The material point method for simulating continuum materials},
  author={Jiang, Chenfanfu and Schroeder, Craig and Teran, Joseph},
  journal={Communications of the ACM},
  volume={61},
  number={3},
  pages={86--95},
  year={2016},
  publisher={ACM}
}

@article{rombach2022high,
  title={High-resolution image synthesis with latent diffusion models},
  author={Rombach, Robin and Blattmann, Andreas and Lorenz, Dominik and Esser, Patrick and Ommer, Bj{\"o}rn},
  journal={Proceedings of the IEEE/CVF Conference on Computer Vision and Pattern Recognition},
  pages={10684--10695},
  year={2022}
}

@article{hu2018moving,
  title={A moving least squares material point method with displacement discontinuity and two-way rigid body coupling},
  author={Hu, Yuanming and Fang, Yu and Ge, Ziheng and Qu, Ziyin and Zhu, Yixin and Pradhana, Andre and Jiang, Chenfanfu},
  journal={ACM Transactions on Graphics (TOG)},
  volume={37},
  number={4},
  pages={1--14},
  year={2018},
  publisher={ACM New York, NY, USA}
}

@inproceedings{xie2024physgaussian,
  title={Physgaussian: Physics-integrated 3d gaussians for generative dynamics},
  author={Xie, Tianyi and Zong, Zeshun and Qiu, Yuxing and Li, Xuan and Feng, Yutao and Yang, Yin and Jiang, Chenfanfu},
  booktitle={Proceedings of the IEEE/CVF Conference on Computer Vision and Pattern Recognition},
  pages={4389--4398},
  year={2024}
}

@article{liu2024physics3d,
  title={Physics3d: Learning physical properties of 3d gaussians via video diffusion},
  author={Liu, Fangfu and Wang, Hanyang and Yao, Shunyu and Zhang, Shengjun and Zhou, Jie and Duan, Yueqi},
  journal={arXiv preprint arXiv:2406.04338},
  year={2024}
}

@article{lin2025omniphysgs,
  title={OmniPhysGS: 3D Constitutive Gaussians for General Physics-Based Dynamics Generation},
  author={Lin, Yuchen and Lin, Chenguo and Xu, Jianjin and Mu, Yadong},
  journal={arXiv preprint arXiv:2501.18982},
  year={2025}
}

@inproceedings{zhang2024physdreamer,
  title={Physdreamer: Physics-based interaction with 3d objects via video generation},
  author={Zhang, Tianyuan and Yu, Hong-Xing and Wu, Rundi and Feng, Brandon Y and Zheng, Changxi and Snavely, Noah and Wu, Jiajun and Freeman, William T},
  booktitle={European Conference on Computer Vision},
  pages={388--406},
  year={2024},
  organization={Springer}
}

@misc{senseglove,
  author       = {{SENSEGLOVE}},
  title        = {SenseGlove},
  howpublished = {\url{https://www.senseglove.com/}},
  year         = {2025},
  note         = {Accessed: 2025-07-11}
}

@inproceedings{wimmer2025gaussians,
  title={Gaussians-to-Life: Text-Driven Animation of 3D Gaussian Splatting Scenes},
  author={Wimmer, Thomas and Oechsle, Michael and Niemeyer, Michael and Tombari, Federico},
  booktitle={2025 International Conference on 3D Vision},
  pages={958--968},
  year={2025},
  organization={IEEE}
}

@article{ren2023dreamgaussian4d,
  title={Dreamgaussian4d: Generative 4d gaussian splatting},
  author={Ren, Jiawei and Pan, Liang and Tang, Jiaxiang and Zhang, Chi and Cao, Ang and Zeng, Gang and Liu, Ziwei},
  journal={arXiv preprint arXiv:2312.17142},
  year={2023}
}

@article{bansal2025videophy,
  title={Videophy-2: A challenging action-centric physical commonsense evaluation in video generation},
  author={Bansal, Hritik and Peng, Clark and Bitton, Yonatan and Goldenberg, Roman and Grover, Aditya and Chang, Kai-Wei},
  journal={arXiv preprint arXiv:2503.06800},
  year={2025}
}

@inproceedings{chen2025physgen3d,
  title={Physgen3d: Crafting a miniature interactive world from a single image},
  author={Chen, Boyuan and Jiang, Hanxiao and Liu, Shaowei and Gupta, Saurabh and Li, Yunzhu and Zhao, Hao and Wang, Shenlong},
  booktitle={Proceedings of the Computer Vision and Pattern Recognition Conference},
  pages={6178--6189},
  year={2025}
}

@inproceedings{huang2024vbench,
  title={Vbench: Comprehensive benchmark suite for video generative models},
  author={Huang, Ziqi and He, Yinan and Yu, Jiashuo and Zhang, Fan and Si, Chenyang and Jiang, Yuming and Zhang, Yuanhan and Wu, Tianxing and Jin, Qingyang and Chanpaisit, Nattapol and others},
  booktitle={Proceedings of the IEEE/CVF Conference on Computer Vision and Pattern Recognition},
  pages={21807--21818},
  year={2024}
}

@article{guo2023animatediff,
  title={AnimateDiff: Animate Your Personalized Text-to-Image Diffusion Models without Specific Tuning},
  author={Yuwei Guo and Ceyuan Yang and Anyi Rao and Zhengyang Liang and Yaohui Wang and Yu Qiao and Maneesh Agrawala and Dahua Lin and Bo Dai},
  journal={arXiv preprint arxiv:2307.04725},
  year={2023},
  archivePrefix={arXiv},
  primaryClass={cs.CV}
}

@article{yang2024cogvideox,
  title={Cogvideox: Text-to-video diffusion models with an expert transformer},
  author={Yang, Zhuoyi and Teng, Jiayan and Zheng, Wendi and Ding, Ming and Huang, Shiyu and Xu, Jiazheng and Yang, Yuanming and Hong, Wenyi and Zhang, Xiaohan and Feng, Guanyu and others},
  journal={arXiv preprint arXiv:2408.06072},
  year={2024}
}

@inproceedings{xing2024dynamicrafter,
  title={Dynamicrafter: Animating open-domain images with video diffusion priors},
  author={Xing, Jinbo and Xia, Menghan and Zhang, Yong and Chen, Haoxin and Yu, Wangbo and Liu, Hanyuan and Liu, Gongye and Wang, Xintao and Shan, Ying and Wong, Tien-Tsin},
  booktitle={European Conference on Computer Vision},
  pages={399--417},
  year={2024},
  organization={Springer}
}

@article{zheng2024open,
  title={Open-sora: Democratizing efficient video production for all},
  author={Zheng, Zangwei and Peng, Xiangyu and Yang, Tianji and Shen, Chenhui and Li, Shenggui and Liu, Hongxin and Zhou, Yukun and Li, Tianyi and You, Yang},
  journal={arXiv preprint arXiv:2412.20404},
  year={2024}
}

@inproceedings{chang2019d3tw,
  title={D3tw: Discriminative differentiable dynamic time warping for weakly supervised action alignment and segmentation},
  author={Chang, Chien-Yi and Huang, De-An and Sui, Yanan and Fei-Fei, Li and Niebles, Juan Carlos},
  booktitle={Proceedings of the IEEE/CVF Conference on Computer Vision and Pattern Recognition},
  pages={3546--3555},
  year={2019}
}

@inproceedings{brahmbhatt2020contactpose,
  title={ContactPose: A dataset of grasps with object contact and hand pose},
  author={Brahmbhatt, Samarth and Tang, Chengcheng and Twigg, Christopher D and Kemp, Charles C and Hays, James},
  booktitle={European Conference on Computer Vision},
  pages={361--378},
  year={2020},
  organization={Springer}
}

@inproceedings{tse2022s,
  title={S 2 contact: Graph-based network for 3d hand-object contact estimation with semi-supervised learning},
  author={Tse, Tze Ho Elden and Zhang, Zhongqun and Kim, Kwang In and Leonardis, Ales and Zheng, Feng and Chang, Hyung Jin},
  booktitle={European Conference on Computer Vision},
  pages={568--584},
  year={2022},
  organization={Springer}
}

@article{kerbl20233d,
  title={3D Gaussian splatting for real-time radiance field rendering.},
  author={Kerbl, Bernhard and Kopanas, Georgios and Leimk{\"u}hler, Thomas and Drettakis, George},
  journal={ACM Transactions on Graphics (TOG)},
  volume={42},
  number={4},
  pages={139--1},
  year={2023}
}

@inproceedings{wu20244d,
  title={4D gaussian splatting for real-time dynamic scene rendering},
  author={Wu, Guanjun and Yi, Taoran and Fang, Jiemin and Xie, Lingxi and Zhang, Xiaopeng and Wei, Wei and Liu, Wenyu and Tian, Qi and Wang, Xinggang},
  booktitle={Proceedings of the IEEE/CVF Conference on Computer Vision and Pattern Recognition},
  pages={20310--20320},
  year={2024}
}

@inproceedings{jiang2024vr,
  title={Vr-gs: A physical dynamics-aware interactive gaussian splatting system in virtual reality},
  author={Jiang, Ying and Yu, Chang and Xie, Tianyi and Li, Xuan and Feng, Yutao and Wang, Huamin and Li, Minchen and Lau, Henry and Gao, Feng and Yang, Yin and others},
  booktitle={ACM SIGGRAPH 2024 Conference Papers},
  pages={1--1},
  year={2024}
}

@inproceedings{zhang2023adding,
  title={Adding conditional control to text-to-image diffusion models},
  author={Zhang, Lvmin and Rao, Anyi and Agrawala, Maneesh},
  booktitle={Proceedings of the IEEE/CVF International Conference on Computer Vision},
  pages={3836--3847},
  year={2023}
}

@inproceedings{huang2025dreamphysics,
  title={DreamPhysics: Learning Physics-Based 3D Dynamics with Video Diffusion Priors},
  author={Huang, Tianyu and Zhang, Haoze and Zeng, Yihan and Zhang, Zhilu and Li, Hui and Zuo, Wangmeng and Lau, Rynson WH},
  booktitle={Proceedings of the AAAI Conference on Artificial Intelligence},
  volume={39},
  pages={3733--3741},
  year={2025}
}

@article{ni2024phyrecon,
  title={Phyrecon: Physically plausible neural scene reconstruction},
  author={Ni, Junfeng and Chen, Yixin and Jing, Bohan and Jiang, Nan and Wang, Bin and Dai, Bo and Li, Puhao and Zhu, Yixin and Zhu, Song-Chun and Huang, Siyuan},
  journal={Advances in Neural Information Processing Systems},
  volume={37},
  pages={25747--25780},
  year={2024}
}

@article{yu20244real,
  title={4real: Towards photorealistic 4d scene generation via video diffusion models},
  author={Yu, Heng and Wang, Chaoyang and Zhuang, Peiye and Menapace, Willi and Siarohin, Aliaksandr and Cao, Junli and Jeni, L{\'a}szl{\'o} and Tulyakov, Sergey and Lee, Hsin-Ying},
  journal={Advances in Neural Information Processing Systems},
  volume={37},
  pages={45256--45280},
  year={2024}
}

@article{zhuang2024tip,
  title={Tip-editor: An accurate 3d editor following both text-prompts and image-prompts},
  author={Zhuang, Jingyu and Kang, Di and Cao, Yan-Pei and Li, Guanbin and Lin, Liang and Shan, Ying},
  journal={ACM Transactions on Graphics (TOG)},
  volume={43},
  number={4},
  pages={1--12},
  year={2024},
  publisher={ACM New York, NY, USA}
}

@inproceedings{li2024animatable,
  title={Animatable gaussians: Learning pose-dependent gaussian maps for high-fidelity human avatar modeling},
  author={Li, Zhe and Zheng, Zerong and Wang, Lizhen and Liu, Yebin},
  booktitle={Proceedings of the IEEE/CVF Conference on Computer Vision and Pattern Recognition},
  pages={19711--19722},
  year={2024}
}

@inproceedings{wu2024gaussctrl,
  title={Gaussctrl: Multi-view consistent text-driven 3d gaussian splatting editing},
  author={Wu, Jing and Bian, Jia-Wang and Li, Xinghui and Wang, Guangrun and Reid, Ian and Torr, Philip and Prisacariu, Victor Adrian},
  booktitle={European Conference on Computer Vision},
  pages={55--71},
  year={2024},
  organization={Springer}
}

@article{lin2025diffsplat,
  title={Diffsplat: Repurposing image diffusion models for scalable gaussian splat generation},
  author={Lin, Chenguo and Pan, Panwang and Yang, Bangbang and Li, Zeming and Mu, Yadong},
  journal={arXiv preprint arXiv:2501.16764},
  year={2025}
}

@article{gomel2024diffusion,
  title={Diffusion-Based Attention Warping for Consistent 3D Scene Editing},
  author={Gomel, Eyal and Wolf, Lior},
  journal={arXiv preprint arXiv:2412.07984},
  year={2024}
}

@article{tang2023dreamgaussian,
  title={Dreamgaussian: Generative gaussian splatting for efficient 3d content creation},
  author={Tang, Jiaxiang and Ren, Jiawei and Zhou, Hang and Liu, Ziwei and Zeng, Gang},
  journal={arXiv preprint arXiv:2309.16653},
  year={2023}
}

@article{howil2025clipgaussian,
  title={CLIPGaussian: Universal and Multimodal Style Transfer Based on Gaussian Splatting},
  author={Howil, Kornel and Borycki, Piotr and Dziarmaga, Tadeusz and Mazur, Marcin and Spurek, Przemys{\'L} and others},
  journal={arXiv preprint arXiv:2505.22854},
  year={2025}
}

@article{liu2025abc,
  title={ABC-GS: Alignment-Based Controllable Style Transfer for 3D Gaussian Splatting},
  author={Liu, Wenjie and Liu, Zhongliang and Yang, Xiaoyan and Sha, Man and Li, Yang},
  journal={arXiv preprint arXiv:2503.22218},
  year={2025}
}

@misc{BlenderNeRF,
  author       = {Raafat, Maxime},
  title        = {BlenderNeRF},
  year         = {2024},
  url          = {https://github.com/maximeraafat/BlenderNeRF},
  note         = {Accessed: 2025-09-07}
}

@article{bansal2024videophy,
  title={Videophy: Evaluating physical commonsense for video generation},
  author={Bansal, Hritik and Lin, Zongyu and Xie, Tianyi and Zong, Zeshun and Yarom, Michal and Bitton, Yonatan and Jiang, Chenfanfu and Sun, Yizhou and Chang, Kai-Wei and Grover, Aditya},
  journal={arXiv preprint arXiv:2406.03520},
  year={2024}
}

@INPROCEEDINGS{601047,
  author={Moccozet, L. and Thalmann, N.M.},
  booktitle={Proceedings. Computer Animation '97 (Cat. No.97TB100120)}, 
  title={Dirichlet free-form deformations and their application to hand simulation}, 
  year={1997},
  volume={},
  number={},
  pages={93-102},
  doi={10.1109/CA.1997.601047}}

@article{hartmann2022method,
  title={A method to estimate contact regions between hands and objects during human multi-digit grasping},
  author={Hartmann, Frieder and Maiello, Guido and Rothkopf, Constantin A and Fleming, Roland W},
  journal={bioRxiv},
  pages={2022--09},
  year={2022},
  publisher={Cold Spring Harbor Laboratory}
}

@inproceedings{chen2024gaussianeditor,
  title={Gaussianeditor: Swift and controllable 3d editing with gaussian splatting},
  author={Chen, Yiwen and Chen, Zilong and Zhang, Chi and Wang, Feng and Yang, Xiaofeng and Wang, Yikai and Cai, Zhongang and Yang, Lei and Liu, Huaping and Lin, Guosheng},
  booktitle={Proceedings of the IEEE/CVF Conference on Computer Vision and Pattern Recognition},
  pages={21476--21485},
  year={2024}
}

@inproceedings{zhou2024feature,
  title={Feature 3dgs: Supercharging 3d gaussian splatting to enable distilled feature fields},
  author={Zhou, Shijie and Chang, Haoran and Jiang, Sicheng and Fan, Zhiwen and Zhu, Zehao and Xu, Dejia and Chari, Pradyumna and You, Suya and Wang, Zhangyang and Kadambi, Achuta},
  booktitle={Proceedings of the IEEE/CVF Conference on Computer Vision and Pattern Recognition},
  pages={21676--21685},
  year={2024}
}

@inproceedings{huang2024sc,
  title={Sc-gs: Sparse-controlled gaussian splatting for editable dynamic scenes},
  author={Huang, Yi-Hua and Sun, Yang-Tian and Yang, Ziyi and Lyu, Xiaoyang and Cao, Yan-Pei and Qi, Xiaojuan},
  booktitle={Proceedings of the IEEE/CVF Conference on Computer Vision and Pattern Recognition},
  pages={4220--4230},
  year={2024}
}

@inproceedings{guedon2024sugar,
  title={Sugar: Surface-aligned gaussian splatting for efficient 3d mesh reconstruction and high-quality mesh rendering},
  author={Gu{\'e}don, Antoine and Lepetit, Vincent},
  booktitle={Proceedings of the IEEE/CVF Conference on Computer Vision and Pattern Recognition},
  pages={5354--5363},
  year={2024}
}

@article{chen2024gi,
  title={Gi-gs: Global illumination decomposition on gaussian splatting for inverse rendering},
  author={Chen, Hongze and Lin, Zehong and Zhang, Jun},
  journal={arXiv preprint arXiv:2410.02619},
  year={2024}
}

@inproceedings{feng2024pie,
  title={Pie-nerf: Physics-based interactive elastodynamics with nerf},
  author={Feng, Yutao and Shang, Yintong and Li, Xuan and Shao, Tianjia and Jiang, Chenfanfu and Yang, Yin},
  booktitle={Proceedings of the IEEE/CVF Conference on Computer Vision and Pattern Recognition},
  pages={4450--4461},
  year={2024}
}

@InProceedings{Kayan_2025_ICCV,
    author    = {Kayan, Karhan and Alexandropoulos, Stamatis and Jain, Rishabh and Zuo, Yiming and Liang, Erich and Deng, Jia},
    title     = {Princeton365: A Diverse Dataset with Accurate Camera Pose},
    booktitle = {Proceedings of the IEEE/CVF International Conference on Computer Vision},
    month     = {October},
    year      = {2025},
    pages     = {7645-7654}
}

@article{liu2024frechetvideomotiondistance,
  title={Fr{\'e}chet video motion distance: a metric for evaluating motion consistency in videos },
  author={Liu, Jiahe and Qu, Youran and Yan, Qi and Zeng, Xiaohui and Wang, Lele and Liao, Renjie},
  journal={arXiv preprint arXiv:2407.16124},
  year={2024}
}
\end{document}